\documentclass{cjaa-1}                   
\usepackage{graphicx}                   
\input{epsf.sty}                        
\input{psfig.sty}                       

\begin{document}

   \title{Polarization in Gamma-Ray Bursts Produced by Pinch Discharge
}

   \volnopage{Vol.0 (200x) No.0, 000--000}      
   \setcounter{page}{1}          

\author{Mei Wu
      \inst{1}
   \and Li Chen
      \inst{2}
   \and Ti-Pei Li
      \inst{1,3} 
      }
\email{chenli@bnu.edu.cn}
   \offprints{T.P. Li}                   

\institute{
Key Lab. of Particle Astrophys., Inst. of High Energy Phys., Chinese Academy of
Sciences\\
 \email{litp@mail.tsinghua.edu.cn}
   \and
Department of Astronomy, Beijing Normal University
   \and
Center for Astrophysics, Tsinghua
University, Beijing
}

   \date{ }

   \abstract{Large-voltage and high-temperature plasma columns produced by pinch
discharge can generate
$\gamma$-ray flashes with energy spectra and spectral evolution consistent
with that observed in $\gamma$-ray bursts (GRBs).
The inverse Compton scattering (ICS) during the
discharge process can produce high linear polarization.
The calculation indicates that the observed polarization
depends on the angle between the observer's line-of-sight
to the GRB and the direction of the pinch discharge, but only weakly depends
on observed $\gamma$-ray energy.
 }

   \authorrunning{M. Wu, T. P. Li \& L. Chen}            
   \titlerunning{Polarization in GRBs Produced by Pinch Discharge}  

   \maketitle

%
%
\section{INTRODUCTION}
Coburn and Boggs (2003) reported their analysis of GRB 021206
observed by the Reuven Ramaty High Energy Solar Spectroscopic
Imager (RHESSI). They found that a high degree,  $(80\pm20)\%$, of
linear polarization is present in the prompt emission in 25-2000
keV. But, due to the limited ratio of signal to noise of the data,
Rutledge \& Fox (2004) concluded a null polarization
and Wigger et al. (2004) obtained the linear
polarization degree of $(41^{+57}_{-44})\%$ from the same data.
The study of polarization is important in understanding the production
mechanism of GRBs, though the detection is still uncertain
at the present stage. It is expected that more reliable measurements
in the future for GRB polarization may give strong constraint
on different GRB models.

Based on the fireball model, some theoretical considerations
have been done to explain the possible polarization. Lyutikov et
al. (2003) believed that the large scale ordered magnetic fields
produced at the central source may produce highly polarized
GRB prompt emission.
The calculation demanded the magnetic field in the emission region
is dominated by the toroidal field and is concentrated in a thin
shell near the surface of the shell expanding with the Lorentz
factor and the synchrotron emission is produced by an isotropic
population of relativistic electrons with the power law
distribution in energy. Strong
polarization might arise in a jet with the line-of-sight to the GRB
being close to the jet edge (Medvedev \& Loeb 1999; Gruzinov 1999; 
Waxman 2003). In this case, the polarization signal may
be not averaged out. Dar \& De Rujula (2003) and Eichler \&
Levinson (2003) discussed polarization as a characteristic
signature of inverse Compton up-scatter. Lazzati et al. (2003)
considered Compton drag as a mechanism for high linear
polarization. In such scenario, the relativistic electrons are
contained in a fireball, and the shape of the soft photon field is
selected in order to reproduce the GRB spectra.

We have proposed an electrical discharge model of GRBs (Li \& Wu 1997):
large-voltage and high-temperature pinch plasma columns
produced by disruptive electrical discharges can generate $\gamma$-ray
flashes with energy spectra and spectral evolution characters consistent
with that observed in GRBs.
In this paper we show that the pinch discharge process can naturally
produce the high degree polarization.

\section{DISCHARGE MODEL}
\begin{figure}
\vspace{-2.5cm}
 \includegraphics[width=120mm]{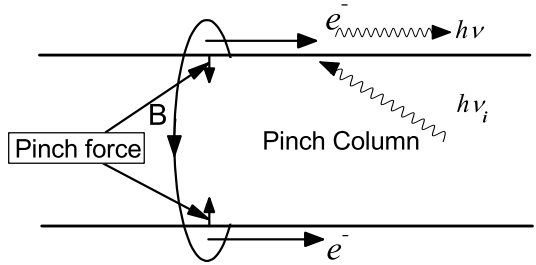}
\vspace{-2.5cm}
 \caption{The pinch discharge model of GRBs.
\label{f:pd}}
  \end{figure}

Alfv\'{e}n (1981) stressed the importance of electric currents
in plasmas to understanding phenomena occurred in the
magnetosphere till galactic dimensions.  A current flowing in the plasma
may contract by the magnetic confinement and form a plasma cable with much
larger density than the surroundings. He held that many of the explosive
events observed in cosmic physics are produced by disruptive discharges
of electric double layers in the current cables.
The disruptive discharge will cause the plasma cable to pinch
into a very narrow column by inward magnetic pressure
of the discharge current  (Krall \& Trivelpiece 1973).
The resulted compressed, large-voltage and high-temperature discharge
column will radiate flaring energetic photons due to ICS
between the accelerated electrons and thermal photons, as sketched in Fig.~\ref{f:pd}.

In an isotropic thermal emission with mean photon
energy $\overline{h\nu}$, the mean energy of a
scattered photon from an electron with energy $\epsilon=\gamma mc^2$ is
$\bar{E}=\frac{4}{3}\gamma^2\overline{h\nu}$
(Ginzburg \& Syrovatskii 1964).
For a pinch discharge column with temperature $T=10^6$ K,
the mean photon energy $\overline{h\nu}\sim 230$ eV,
then the mean energy of the scattered photons from electrons with $\gamma=30$
($\epsilon\sim 15$ MeV) will be $\bar{E}\sim 276$ keV.
For a blackbody emission of $10^6$ K, the photon density $n\approx 2\times 10^{19}/$cm$^3$,
the Compton cross-section $\sigma_c\approx 6\times 10^{-25}$ cm$^2$,
then the electron free path $\l_c=1/n\sigma_c\approx 1$ km.
A large voltage and high temperature discharge column with a length
$l \ga l_c$ should be an efficient radiator of hard X-rays and soft $\gamma$-rays.

The TGFs, intense gamma-ray flashes of atmospheric origin observed by BATSE/$CGRO$,
are quite similar with GRBs in emission energy band and morphology.
An evident correlation of TGFs with thunderstorm system (Fishman et al. 1994) and
their features of energy dependence of time profiles (Feng et al. 2002) indicate
that TGFs are produced by upward explosive electrical discharges at high altitude,
giving an observational evidence that the discharge process can produce high-energy
explosive events.
 Before a burst the energy is supposed to be stored in an electric circuit with a current $I$
and inductance $L$ as the inductive energy  $W=L I^2/2$.
It has been realized that there is an atmospheric electrical global circuit,
in which a current flows between the ionosphere and the earth
with thunderstorms as a d.c. generator (see e.g. Markson \& Muir 1980).
Another kind of high-energy transients similar to GRBs is the
solar hard X-ray flare. Alfv\'{e}n \& Carlqvist (1967) suggested exploding discharges
of electric double layers to be responsible for solar flares.
In their model, the energy release and particle acceleration of a flare
are produced by the disruption discharge of a current of $10^{11}-10^{12}$ A, flowing in a solar
atmosphere circuit with a typical length $10^9-10^{10}$ cm and inductance $\sim 10$ H.

The pinch discharge scenario could account
for the energy budget even for GRBs at cosmological distances.
For the typical $10^{53}$ erg isotropic-equivalent output,
collimation reduces the burst energy budget to $\sim 10^{49}$ erg.
For a circuit with a radius of stellar dimension, we can expect its
inductance $L\sim 10^2$ H, then the needed current $I\sim 10^{20}$ A.
The required current circuit is not impossible in the astrophysical context (see e.g.,
Raadu 1989).
Large-scale current circuits have been proposed both for the heliosphere and for
the galaxy (Alfv\'{e}n 1978). In the galactic case currents of $10^{17}-10^{19}$~A
have been estimated (Alfv\'{e}n \& Carlqvist 1978).
Since the typical GRB photon energy is
$\sim 100$ keV, $\sim 10^{56}$ photons must be scattered to lead to a GRB.
The needed radius $r$ and length $l$ of the discharge column
to produce $N_b=10^{56}$ photons in $t_{burst}=30$ s can be estimated by $2\pi rlft_{burst}=N_b$,
where $f$ is the emission intensity of black body with temperature $kT$ (keV),
$f= 5\times 10^{32} (kT)^3$ cm$^{-2}$ s$^{-1}$. Then the required column size
$rl\sim 10^{21}/(kT)^3$ cm$^2$. Bursts consisting of spiky components are produced
by branching discharges, the size of each individual column can be much smaller.

The pinch discharge mechanism can naturally interpret many observed GRB
characteristics (Li \& Wu 1997; Li 1998). Complicated morphological patterns in GRBs,
such as wide variety of profile configurations, rich in fluctuation or smooth
structures, rapid rise vs slower fall, weak precursor and secondary pulse,
etc., are common in various electrical discharges.
Producing energy spectra of smoothly-joining broken power law,
so called the Band model to describe GRB spectra, is a general property of
the discharge mechanism.
The distribution feature of the low-energy power law slope $\alpha$ and
peak energy $E_{peak}$ from the discharge model is similar to what
observed in GRBs. Spectral evolution features observed in GRBs,
e.g. hard-to-soft spectral evolution and time-resolved low-energy
spectra  over a pulse, can be reproduced by the discharge model.
The typical energy dependence of GRB time profiles, the variability
of hard emission is earlier and narrower than that of soft emission,
is the character of a disruptive discharge mechanism as well.
Here we will show in the next section that the discharge mechanism
can also naturally produce strongly polarized emission.

\section{POLARIZATION}
In a pinch discharge in plasma, the electric current is along the surface
of the discharge column by the skin effect. The high-temperature pinched
plasma is compressed within the column.
In the following calculations we suppose the local surface of a concerned pinch
column can be seen as a plane, then the electric lines of force on it
are parallel to each other.

\begin{figure}
\vspace{-2.5cm}
 \includegraphics[width=120mm]{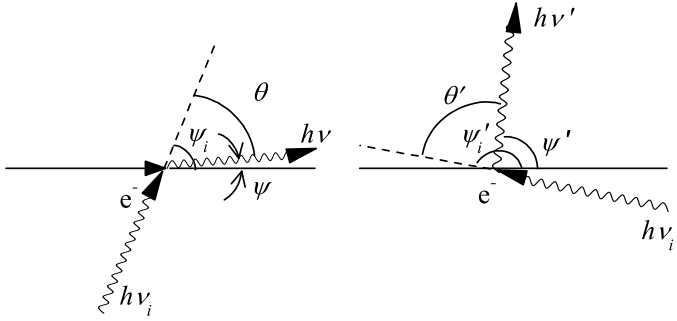}
\vspace{-2.5cm}
 \caption{Geometry for inverse Compton scattering.
{\it Left panel}: in the lab system. {\it Right panel}:
in the electron rest system. \label{f:compt}
}
  \end{figure}

Fig.~\ref{f:compt} illustrates the ICS process of a photon with incident energy $h\nu_i$
and a moving electron to produce a scattered photon with energy $h\nu$,
where $\psi_i$ and $\psi$ are the
include angle of the incident and scattered photon with respect to
the moving electron respectively, and the scattering angle between
the incident and scattered photons $\theta=\psi_i-\psi$. With Lorentz
transformations, the energy ratio between the incident and scattered photons
in the electron rest frame (in primed notation) can be derived as
\begin{eqnarray}
\frac{h\nu_i^{\prime}}{h\nu^{\prime}}&=&1+\frac{\gamma
h\nu_i}{mc^2}(1-\beta \cos\psi_i) \equiv
\lambda(\psi^{\prime},\theta^{\prime},h\nu_i)~,
\end{eqnarray}
where $\beta$ is the electron velocity in units of $c$,
$\gamma=\sqrt{1-\beta^2}$, and $m$ the electron rest mass. We also
used the relations $\psi_i=\arccos\left(
\frac{\beta+\cos\psi_i^{\prime}}{1+\beta\cos\psi_i^{\prime}}\right)
$ and $\psi_i^{\prime}=\theta^{\prime}+\psi^{\prime}$.

At the scattering angle $\theta^{\prime}$ the polarization $\xi$
of an ICS photon with incident energy $h\nu_i^{\prime}$
 can be expressed as Akhiezer \& Berestetskii (1965).
\begin{equation}
\xi(\psi^{\prime},\theta^{\prime},h\nu_i )=
\frac{\sin^2\theta^{\prime}}{\frac{h\nu^{\prime}_i}{h\nu^{\prime}}+
\frac{h\nu^{\prime}}{h\nu^{\prime}_i}-\sin^2\theta^{\prime}}
=\frac{\sin^2\theta^{\prime}}{\lambda+\lambda^{-1}-\sin^2\theta^{\prime}}~.
\end{equation}
The differential scattering cross-section in the electron rest
system is (Klein  \& Nishina 1929)
\begin{eqnarray}
\frac{d\sigma_{T}}{d\psi^{\prime}}
 & \propto & \left(\frac{h\nu^{\prime}}{h\nu_i^{\prime}}\right)^2
\left(\frac{h\nu^{\prime}_i}{h\nu^{\prime}}+\frac{h\nu^{\prime}}
{h\nu^{\prime}_i}-\sin^2\theta^{\prime}\right)\sin\psi^{\prime}  \nonumber\\
& = & \lambda^{-2}(\lambda+\lambda^{-1}-\sin^2\theta^{\prime})
\sin\psi^{\prime} \nonumber \\
& \equiv & \mu(\psi^{\prime},\theta^{\prime},h\nu_i)~.
\end{eqnarray}
For isotropic photons in the lab system, the number density of
photons collided with the moving electron at a certain incident
angle $\psi_i$ is
\begin{equation}
 \frac{dN}{d\psi_i}\propto (1-\beta\cos\psi_i)\sin\psi_i
 \equiv\eta(\psi^{\prime},\theta^{\prime})~.
\label{nd}
\end{equation}

The polarization is
a Lorentz invariant, we can calculate it in the electron rest frame.
From ICS of photons with energy spectrum $p(h\nu_i)$ in the lab system,
the average polarization $\bar{\xi}(\psi)$ of scattered photons at a certain
angle $\psi$ can be calculated as
\begin{equation}
\bar{\xi}(\psi)=\bar{\xi}(\psi^{\prime})
=\frac{\int\xi(\psi^{\prime},\theta^{\prime},h\nu_i)\mu(\psi^{\prime},
\theta^{\prime},h\nu_i)\eta(\psi^{\prime},\theta^{\prime})p(h\nu_i)d\theta^{\prime}d(h\nu_i)}
{\int
\mu(\psi^{\prime},\theta^{\prime},h\nu_i)\eta(\psi^{\prime},\theta^{\prime})p(h\nu_i)
d\theta^{\prime}d(h\nu_i) }~.
\end{equation}

\begin{figure}
 \includegraphics[width=120mm]{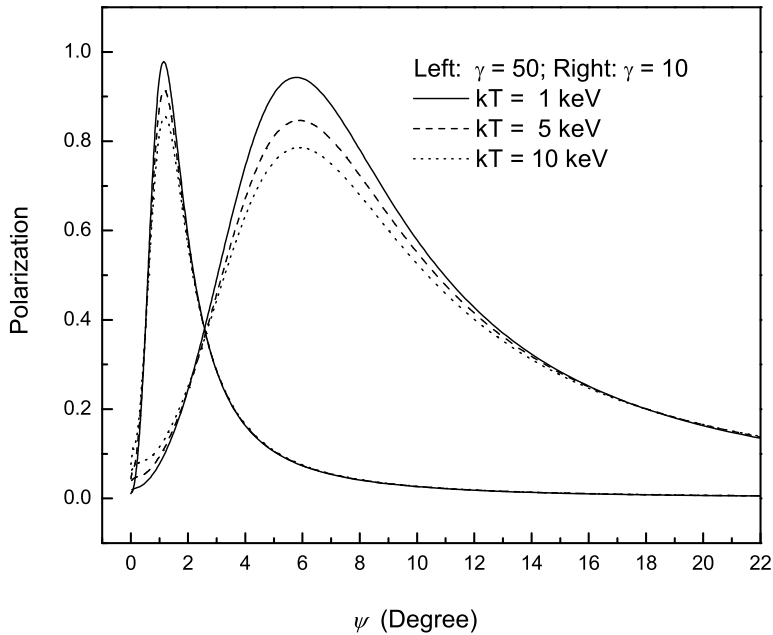}
\vspace{-0.5cm}
 \caption{Polarization vs.
angle $\psi$ of observer's line-of-sight to discharge direction
for electron Lorentz factor $\gamma$=10, 50, and pinched plasma
temperature $kT$=1, 5, 10 keV, respectively. The
calculation is for energy of scattered photons limited in 25-2500
keV. \label{f:pol-ang} }
  \end{figure}

 We calculate the polarization at different angles $\psi$ of the observer's
line-of-sight to the GRB with the discharge direction for photon energy distribution
$p(h\nu_i)$ in the lab system being a blackbody spectrum with $kT=1$, 5 and 10 keV,
separately. For each temperature, the Lorentz factor $\gamma$ of moving electron is taken
to be 10 and 50, respectively. As the pinched high-temperature plasma is
concentrated in the discharge column, incident angles $\psi$  of photon
are restricted in the range of 0 - $\pi$. The results are shown in Fig. \ref{f:pol-ang}.

To see the relationship of polarization to GRB photon energy expected by
the discharge mechanism, we calculate the polarization for different
value of $h\nu$ with given $\gamma,~ kT$ and $\psi$, the results are shown in Fig.~4.

\begin{figure}
 \includegraphics[width=120mm]{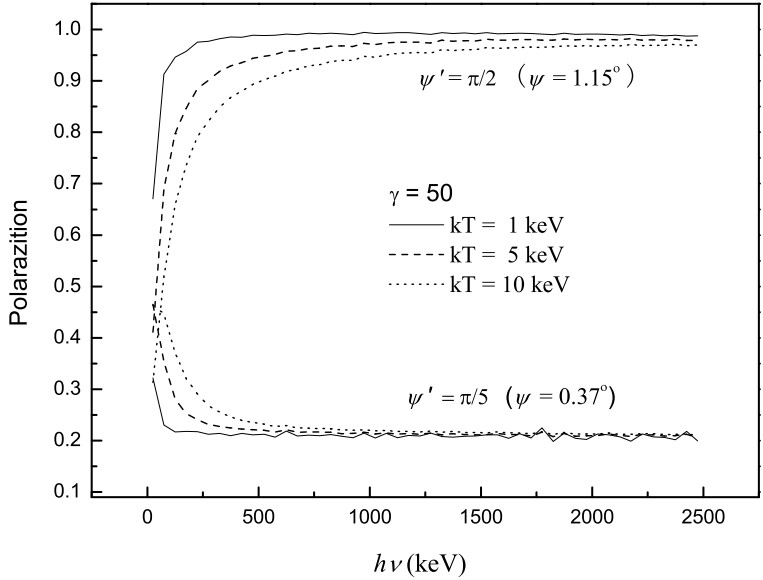}
\vspace{-0.5cm}
 \caption{Polarization vs. energy of
scattered photon for electron Lorentz factor $\gamma=50$
and for different scattered
angles ($\psi^{\prime}=\pi/2,~ \pi/5$)
and pinched plasma temperatures ($kT=1, 5, 10$ keV).
 \label{f:pol-e} }
  \end{figure}

From Figs. 3 and 4 we can see that a discharge column, shown in
Fig.~2, can produce prompt $\gamma$-ray emission polarized in the
direction perpendicular to the scattering plane, the observed
amount of polarization depends on the kinetic energy of the moving
electron, temperature of pinched plasma and angle between the
observer's line-of-sight to the GRB and the direction of the
discharge. It also depends on observed $\gamma$-ray energy, but
for $h\nu\ge 500$ keV the dependency becomes weak.

\section{DISCUSSION}
During a discharge the energy released from the central engine and
after it the pinched high-pressure and high-temperature plasma
may produce fireball and afterglow emission in the ambient medium.
The geometry and environment of the discharge
may influence the production of fireball and afterglow.
Different from fireball shock models, the location of the
GRB prompt emission in the discharge model is neither the internal
shock nor external shock, but the central engine itself.
 In comparison with the blast-wave model,
the discharge mechanism provides a more clear postulate on the nature of the GRB prompt gamma-ray
emission. The GRB producing mechanism in the scenario has no difficulty on the baryon contamination
and compactness problem, and no difficulty on synchrotron
``line-of-death'' (Preece et al. 1998) and cooling problem (Ghisellini et al.
2000) in the synchrotron model.
No need for assuming a globally organized strong magnetic field or
other unusual configuration or bulk motion for fireballs,
the pinch discharge itself can provide necessary conditions
to generate strongly polarized $\gamma$-ray emission: the high-energy electrons
and dense thermal photons for ICS and the discharge column as a preferential direction.

\begin{figure}
\includegraphics[width=120mm]{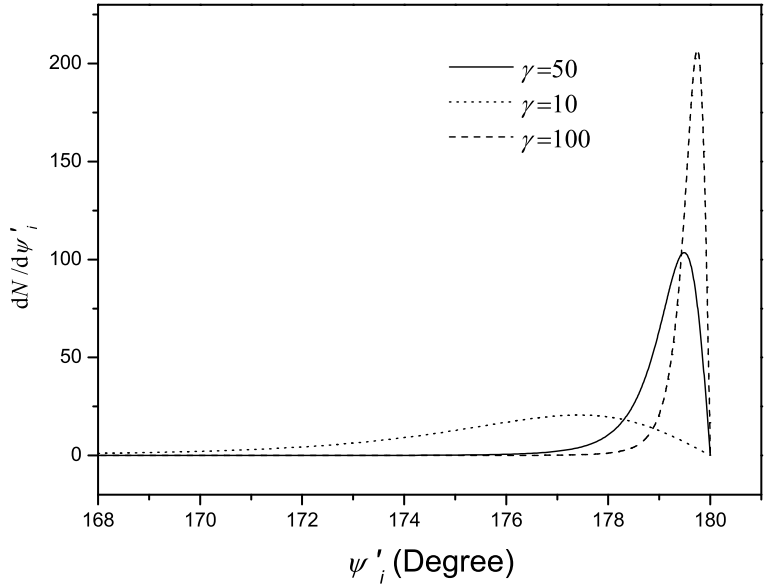}
\vspace{-0.5cm}
 \caption{Distribution of incident angles in the electron
rest frame for isotropic photons in the lab frame.\label{f:nd}
}
\end{figure}

From Eq. (2) one can see that in the electron rest frame, at the
scattered direction perpendicular to the incident one,
$\theta^{\prime}=\pi/2$, the polarization $\xi\simeq 1$ on
condition that \( h\nu^{\prime}\simeq h\nu^{\prime}_i\ll mc^2~. \)
The above condition is usually satisfied by the thermal photons in
a pinch discharge column. For isotropic photons in the lab frame,
we calculate the number density $\frac{dN}{d\psi_i^{\prime}}$ at
an incident angle $\psi_i^{\prime}$ in the electron rest frame
from Eq. (\ref{nd})[noted that $\psi_i=\arccos\left(
\frac{\beta+\cos\psi_i^{\prime}}{1+\beta\cos\psi_i^{\prime}}\right)]
~$with Lorentz transformations for $\gamma=10, 50,$ and 100
respectively and show the result in Fig. (\ref{f:nd}). One can see
that, in the electron rest frame, most photons colliding with the
electron concentrate in a tapering half cone with an axis along
the discharge direction. Therefore, high linear polarization can
be observed at a proper observer's angle with the discharge
direction.

 As shown in Fig.~(\ref{f:pol-ang}), the polarization of observed GRBs
depends on the angle $\psi$ between the observer's line-of-sight
and the axis of the pinch column, varying between zero and unit
for a fixed $\gamma$ of discharging electrons. With different
observation angle one can see different polarization (from 0 to
1). The extreme circumstance is null polarization at the position
with $\theta^{\prime}=0$. The angle $\psi_{\textrm{max}}$ where
maximum polarization takes place varies with $\gamma$. Electrons
moving along a discharge column are accelerated by the discharge
voltage and the electron kinetic energy increases along the path.
At a certain moment, the values of $\gamma$ of moving electrons
should be distributed over a wide range, and hence high level of
polarization can be observed over a considerable range, say
several degrees, of observation angle.

\begin{acknowledgements}
The referee and Dr. Lu Fangjun are thanked for
helpful comments and suggestions on the manuscript.
This work is supported by the Special Funds for Major State Basic Research
Projects and the National Natural Science Foundation of China.
\end{acknowledgements}

\end{document}